\documentclass[aps,prb,preprint,nofootinbib]{revtex4-1}
\usepackage{graphicx}
\usepackage[lofdepth,lotdepth]{subfig}

\usepackage{calrsfs}
\usepackage{mathtools}
\usepackage{amssymb}
\usepackage{amsmath}
\usepackage{dsfont}
\usepackage{hyperref}
\usepackage{braket}
\DeclareMathAlphabet{\pazocal}{OMS}{zplm}{m}{n}

\usepackage{lineno}

\numberwithin{equation}{section}

\begin{document}
\title{Majorana Fermions and Orthogonal Complex Structures}
\author{J.S. Calder\'{o}n-Garc\'ia}
\email[]{js.calderon3227@uniandes.edu.co}
\affiliation{Departamento de F\'{i}sica, Universidad de los Andes,  A.A. 4976-12340, Bogot\'a, Colombia}

\author{A. F. Reyes-Lega}
\email[]{anreyes@uniandes.edu.co}
\affiliation{Departamento de F\'{i}sica, Universidad de los Andes,  A.A. 4976-12340, Bogot\'a, Colombia}

\date{\today}

\begin{abstract}
Ground states of  quadratic Hamiltonians for fermionic systems can be characterized in terms of  orthogonal complex structures.
The standard way in which  such Hamiltonians are diagonalized makes use  of a certain ``doubling'' of the Hilbert space.
In this work we show that this redundancy in the Hilbert space can be completely lifted if the relevant orthogonal structure
is taken into account. Such an approach allows for a treatment of Majorana fermions which is both physically  and mathematically transparent. Furthermore, an explicit connection between orthogonal complex structures and the topological $\mathbb Z_2$-invariant is given.
\end{abstract}
\maketitle

\section{Introduction}
\label{sec:intro}

The discovery of a quantized conductivity in the Quantum Hall Effect (QHE)~\cite{Klitzing1980} and its subsequent interpretation in terms of a topological invariant, the TKNN invariant~\cite{Thouless1982}, was (retrospectively)  one of the first examples of topological phases of matter. Other examples, like Haldane's honeycomb model~\cite{Haldane1988} showed that in contrast to the QHE, a quantization phenomenon could be present even if (on the average) the magnetic field vanishes. These fields are the main mechanism behind being a breaking of time-reversal (TR) invariance, which under the presence of inversion symmetry leads to QHE-like states.    In more recent years, the relevance of the spin-orbit interaction was recognized and led to the prediction of topologically non-trivial states~\cite{Kane2005} (these states are TR invariant). A prominent feature  of topological insulators is the bulk-boundary correspondence and related existence of edge states, which are protected in the presence of TR invariance~\cite{Hasan2010a}.
All these discoveries have led to a very general topological band theory that has allowed to classify different phases of quantum matter according to dimensionality and  symmetry class. This includes also the class of topological superconductors, for which particle-hole symmetry plays a role analogous to that of TR symmetry for topological insulators, and where   Majorana zero modes play a fundamental role~\cite{Hasan2010a,Kitaev2001}. Furthermore, the recognition of a dependence of the ground state degeneracy on the topology of space for the Fractional QHE, as well as for chiral spin states~\cite{Wen1990}, eventually led to our  current understanding, according to which different phases of matter cannot always be distinguished in terms of symmetry considerations.
Topology is nowadays recognized to play a fundamental role in our understanding of quantum phases of matter~\cite{Asorey2016}.

In the present work we will show how the introduction of orthogonal complex structures in the description of fermionic systems allows to eliminate the Hilbert space redundancy which is familiar for Hamiltonians in the BdG form. We will also establish a direct link between the description of such systems in terms of Clifford and fermionic algebras. In section \ref{S:2} we review the structures that are more relevant for the description of fermionic systems using complex structures. In section \ref{S:3} we then establish different connections between fermionic, Clifford and self-dual algebras. An explicit description of the $\mathbb Z_2$-topological invariant in terms of complex structures is also given. We conclude with a discussion of the results and provide an outlook on future work.
\section{Fermionic systems and orthogonal complex structures}
\label{S:2}
In the standard formalism of second quantization, a fermionic system is described in terms of creation and annihilation operators obeying the canonical anticommutation relations (CAR)
\begin{equation}
\label{eq:1}
\{a_{i},a_{j}^{\dagger}\} = \delta_{ij}, \;\;
\{a_{i}^{\dagger},a_{j}^{\dagger}\} =\{a_{i},a_{j}\}=0,
\end{equation}
in accordance with the Pauli exclusion principle. These operators act on a fermionic Fock space $\pazocal{F}=\bigwedge^{\raisebox{-0.4ex}{\scriptsize $\bullet$}}\pazocal{H}$, where $(\pazocal{H},\langle \cdot,\cdot \rangle)$ denotes the Hilbert space of 1-particle states. When the number of degrees of freedom of the system is infinite (that is, in the quantum field theory limit), the CAR algebra (\ref{eq:1}) can be realized in many \emph{inequivalent} ways. For this reason, it is convenient to distinguish the algebraic properties defining the CAR relations from any particular realization through a Hilbert space representation.
This is similar to what happens in group theory: A given abstract group will have in general many inequivalent representations.

Given a 1-particle Hilbert space $(\pazocal H,\langle\cdot,\cdot\rangle)$, the corresponding  CAR algebra, denoted $\pazocal A_{\mbox{{\tiny CAR}}}(\pazocal H,\langle\cdot,\cdot\rangle)$, is defined~\cite{Bratteli1997} as an algebra with generators of the form $a(u)$ and $a^\dagger(u)$ (for $u\in \pazocal H$), subject to the relations
\begin{eqnarray}
&\lbrace a(u),a^\dagger(v)\rbrace = \langle u,v\rangle,&\nonumber\\
&\lbrace a(u),a(v)\rbrace=\lbrace a^\dagger(u),a^\dagger(v)\rbrace=0.&\label{CAR}
\end{eqnarray}
The diagonalization of a Hamiltonian that is quadratic in the fermionic operators is usually performed by means of a
Bogoliubov transformation. In the context of abstract CAR algebras, Araki~\cite{Araki1968} developed a formalism that
makes use of a ``doubled'' Hilbert space in order to diagonalize quadratic Hamiltonians describing systems with an
infinite number of degrees of freedom. This formalism has several points in common with the Nambu
approach~\cite{Fujikawa2016}. Nevertheless, there is an alternative approach that makes use of orthogonal complex structures
and which, as shown below, provides a setting that is ideal for discussions about Majorana
fermions. Before considering the physics of that approach, let us turn to a brief account of the main mathematical
facts that we need. For details we refer to \cite{Plymen1994,Gracia-Bond'ia2001}.

Consider a real vector space $V$ with $\mathrm{dim}_{\mathbb{R}}(V)=2n$. Let $g(\cdot,\cdot)$ be a positive,
symmetric bilinear form on $V$. An \textit{orthogonal complex structure} is a real linear operator $J:V\rightarrow V$
such that $J^{2}=-1$ and $g(Ju,Jv)=g(u,v)$ for all $u,v\in V$. The idea is to use $J$ in order to
construct a complexification of $V$, which is different from the ordinary one,
$V^\mathbb{C}=V\otimes_{\mathbb{R}}\mathbb{C}$. We define, then, the \textit{complex vector space} $V_{J}$ as the one
obtained from $V$, but with multiplication by (complex) scalars given by $(\alpha+i\beta)v\coloneqq\alpha v+\beta Jv$
for $v\in V$ and $\alpha,\beta\in\mathbb{R}$. In other words, multiplication by $i$ on $V_J$ is given by $iv\coloneqq
Jv$. If we define an inner product in $V_{J}$ by
\begin{equation}\label{eq:<,>_J}
\braket{u,v}_{J}\coloneqq g(u,v)+ig(Ju,v),
\end{equation}
we obtain
a complex  Hilbert space $(V_{J},\braket{\cdot,\cdot}_{J})$ with complex dimension $n$. The last claim arises from the
fact that if we have an orthonormal basis $\{u_{1},\cdots,u_{n}\}$ for $(V_{J},\braket{\cdot,\cdot}_{J})$ then
$\{u_{1},Ju_{1},\dots,u_{n},Ju_{n}\}$ is an orthonormal basis for $(V,g(\cdot,\cdot))$.

The (complex) Clifford algebra $\mathbb{C}\ell(V)$~\cite{Plymen1994} acts naturally on the exterior algebra $\bigwedge^{\raisebox{-0.4ex}{\scriptsize $\bullet$}} V^{\mathbb C}$, but the resulting representation is not irreducible~\cite{Gracia-Bond'ia2001}. If instead we consider an orthogonal complex structure $J$ on $V$, we obtain an irreducible representation on the Fock space $\pazocal{F}_J(V):=\bigwedge^{\raisebox{-0.4ex}{\scriptsize $\bullet$}} V_J$.
As Clifford and CAR algebras are closely related~\cite{Araki1968,Plymen1994,Gracia-Bond'ia2001,Gracia-Bondia1994}, we also obtain an irreducible representation of the CAR algebra
 $\pazocal A_{\mbox{{\tiny CAR}}}(V_J,\langle\cdot,\cdot\rangle_J)$.
 In this representation, creation and annihilation operators $a_{J}(v)$ and $a_{J}^{\dagger}(v)$ acting on $\pazocal{F}_{J}(V)$ are given by:
\begin{align}
&a_{J}^{\dagger}(v)(u_{1}\wedge\dots\wedge u_{k})=v\wedge u_{1}\wedge\dots\wedge u_{k},\nonumber\\
&a_{J}(v)(u_{1}\wedge\dots\wedge u_{k})=\sum_{j=1}^{k}(-1)^{j-1}\braket{v,u_{j}}_{J}u_{1}\wedge\dots\wedge \hat{u}_j\wedge\dots\wedge u_{k},\label{eq:aj}
\end{align}
for $v\in V$ and $u_1,\ldots,u_k\in V_J$.
These operators satisfy the CAR relations  $\{a_{J}(u),a_{J}^{\dagger}(v)\}=\langle u,v\rangle_{J}$,  $\{a^{\dagger}_{J}(u),a^{\dagger}_{J}(v)\}=\{a_{J}(u),a_{J}(v)\}=0$, and give rise to a representation of the (real) Clifford algebra $C\ell(V)$ on  $\pazocal{F}_{J}(V)$. Explicitly, the Clifford generators are given by
 \begin{equation}\label{eq:Clifford-generators}
 \pi_{J}(v)\coloneqq a_{J}^{\dagger}(v)+a_{J}(v).
\end{equation}
The vacuum in $\pazocal{F}_{J}(V)$ can also be characterized as a gaussian state $\omega_J$ with a two-point function given by
\begin{equation}
\langle 0_J|a_J(u)a_J^\dagger(v)|0_J\rangle\equiv\omega_J(a_J(u)a_J^\dagger(v))= \langle u, v\rangle_J.
\end{equation}
In fact, this representation can be obtained from $\omega_J$ (regarded as an algebraic state, cf.~\cite{Balachandran2013}) through the Gelfand-Naimark-Segal (GNS) construction. A most important fact is the possibility (when $\mathrm{dim}V=\infty$) of having inequivalent representations. A very useful  characterization of  the vacuum state $|0_J\rangle$ in the $J$-induced representation is obtained if we extend all operators from $V$ to $V^\mathbb C$, as explained below.

The Clifford generators (\ref{eq:Clifford-generators}), as well as the creation/annihilation operators (\ref{eq:aj}) can be regarded as real linear maps
from $V$ to $\mathcal L (\pazocal F_J(V))$, the space of bounded linear operators on Fock space. These can be extended to complex linear maps
\begin{equation}\label{eq:complex-linear-extension}
  \tilde\pi_{J},\; \tilde a_{J}, \;\tilde a^\dagger_{J}: V^{\mathbb C}
   \longrightarrow \mathcal L (\pazocal F_J(V)),
\end{equation}
which means that, for $\lambda \in \mathbb C$ and $w$ in $V^\mathbb C$, we have $\tilde a_{J}(\lambda w)=\lambda \tilde a_{J}(w)$, as well as  $\tilde a^{\dagger}_{J}(\lambda w)=\lambda \tilde a^{\dagger}_{J}(w)$. But also notice that, since the complex structure on $\pazocal F_J(V)$ is determined by $J$, we also have (for $v$ in $V$):
\begin{equation}
a^\dagger_J(Jv) = i a^\dagger_J(v),\;\; a_J(Jv) = -i a_J(v).
\end{equation}
The minus sign can be traced back to equations
(\ref{eq:<,>_J}) and (\ref{eq:aj}). Summarizing, we have the following important identities ($v\in V$):
\begin{eqnarray}
\tilde a^\dagger_J (i v)  = i a^\dagger_J (v) \equiv J a^\dagger_J (v),\label{eq:identity1}\\
\tilde a_J (i v)   = i a_J (v) \equiv J a_J (v),\label{eq:identity2}\\
a_J^\dagger (J v)  = i a^\dagger_J (v) \equiv J a_J^\dagger (v),\label{eq:identity3}\\
a_J (J v) = -i a_J (v) \equiv -J a_J (v).\label{eq:identity4}
\end{eqnarray}
The complex structure $J$ can also be linearly extended to an operator acting on $V^\mathbb C$. Given that $J^2=-1$, it is only on this space that we can consider the eigenvalue problem for $J$. In fact, the space $V^\mathbb C$ turns out to be the direct sum of the eigenspaces for $J$, with eigenvalues $\pm i$. More concretely, consider the projection operators in $V^\mathbb C$
\begin{equation}
  P_{\pm J} := \frac{1}{2}(1\mp i J),
\end{equation}
and define $W_{\pm J}: = P_{\pm J} (V^\mathbb C)$.
Denote with $g_{\mathbb{C}}$ the complex linear extension of $g$ to $V^\mathbb{C}$. Then, using $\langle\braket{w,z}\rangle\coloneqq 2 g_{\mathbb C}(\overline{w},z)$ as the inner product for $V^\mathbb{C}$, we obtain $W_{-J}=W_{J}^{\perp}$,
 so that
\begin{equation}
\label{eq:W+W_bar}
V^{\mathbb C}= W_J \oplus W_J^\perp.
\end{equation}
Furthermore, restricting
$\langle\braket{\cdot,\cdot}\rangle$ to $W_J$, we obtain the following unitary isomorphism~\cite{Gracia-Bond'ia2001}:
\begin{equation}
\label{eq:V_J=W_J}
(V_{J},\braket{\cdot,\cdot}_{J})\cong(W_{J},\langle\braket{\cdot,\cdot}\rangle).
\end{equation}
Let now $u$ be a vector in $W_J^\perp$. Then we have $P_{-J}(u)= u$ or, equivalently, $u= v + i Jv$, for some $v$ in $V$. Using the identities (\ref{eq:identity1})-(\ref{eq:identity4}) we then obtain $\tilde a^\dagger_J(u)=0$. This, in turn, implies
$\tilde \pi_J(u) |0_J\rangle =0.$
It can be shown~\cite{Gracia-Bond'ia2001} that the opposite is also true. The resulting ``vacuum condition''
\begin{equation}
\label{eq:vacuum-condition}
\tilde \pi_J(u) |0_J\rangle =0 \; \Longleftrightarrow\; u\in W_J^\perp,
\end{equation}
thus provides a full characterization of the vacuum $|0_J\rangle$.

There are several aspects of the above construction that are quite relevant from a physical point of view. The first observation is that for every choice of complex structure $J$ we obtain a vacuum $|0_J\rangle$. What we really mean by this is that every choice of $J$ gives rise to an irreducible representation of the CAR algebra on a Fock space $\pazocal F_J(V)$. Suppose we start with a given, fixed complex structure $J_0$ and now want to find the spectrum of a quadratic Hamiltonian which is given to us in terms of the corresponding  creation and annihilation operators
$a^{(\dagger)}_{J_0}(v)$. The standard way of solving this problem consists in considering linear combinations of such creation and annihilation operators, in such a way that (i) the CAR relations are preserved and (ii) the Hamiltonian becomes diagonal in the new basis. Given any element $h$ in the orthogonal group $O(V,g)$, we obtain a new orthogonal complex structure, by $J_h:= h J_0 h^{-1}$. Moreover, by the universal property of Clifford algebras~\cite{Plymen1994}, such an $h$ induces an automorphism of the Clifford algebra, which is nothing but a Bogoliubov transformation. As the new complex structure is again orthogonal, condition (i) is automatically satisfied. Since the action of $O(V,g)$ on the space $\pazocal J$ of orthogonal complex structures is transitive, condition (ii) is accomplished once we have found a suitable orthogonal transformation. The vacuum condition (\ref{eq:vacuum-condition}) gives a condition, fulfilled by the extended Clifford generators, in terms of an additional, auxiliary space $W_J^\perp$. But notice that the \emph{whole structure} of the CAR algebra depends only on the Hilbert space
$V_J$, which is unitarily equivalent to $W_J$ and has been obtained from a triple
($V, g, J$). This is behind the apparent ``Hilbert space redundancy'' so often found in the literature. In the following section we show that there is actually no redundance whatsoever. The role played by $J$ in the definition of the topological $\mathbb Z_2$-index will also be discussed.
\section{The self-dual formalism and Hilbert space redundancy}
\label{S:3}
\subsection{Quadratic Hamiltonians and self-dual algebra}
Consider a fermionic Hamiltonian of the form
\begin{equation}
\label{eq:quadratic-H}
 H = \sum_{i,j=1}^N \left[ a_i^\dagger A_{ij}a_j +\frac{1}{2}\left( a_i^\dagger B_{ij}a_j^\dagger -
a_i \overline{B}_{ij} a_j  \right)\right],
\end{equation}
where $A$ is a hermitian matrix, and $B$ a  skew-symmetric one. If $B=0$,   the spectrum of $H$ can be readily found upon diagonalizing $A$. If $U$ is a unitary matrix such that $U A U^{\dagger}$ is diagonal, then the Bogoliubov transformation determined by $c_i=\sum_{i}^N U_{ij} a_j$ brings $H$ to a diagonal form. Thus, in this case, diagonalization of the fermionic quadratic form $H=a^\dagger A\, a$ is tantamount to diagonalization of $A$. If $B\neq 0$, we can still regard $H$ as a quadratic form, but only if we ``mix'' creation and annihilation operators. Using a self-explanatory notation, it is easy to check that (up to a constant term) the Hamiltonian can be written in the form
\begin{equation}
\label{eq:q-form}
  H=\frac{1}{2}(a^\dagger, a)\left(
                               \begin{array}{cc}
                                 A & B \\
                                 -\overline{B} & -\overline{A} \\
                               \end{array}
                             \right)\left(
                                      \begin{array}{c}
                                        a \\
                                        a^\dagger \\
                                      \end{array}
                                    \right)
\end{equation}
Araki's construction of the self-dual CAR algebra~\cite{Araki1968} has been devised as an efficient tool to diagonalize quadratic Hamiltonians like (\ref{eq:q-form}), especially in the case $N\rightarrow\infty$. For simplicity, here we will only consider the case of $N$ finite and merely remark that all the arguments presented remain valid in the quantum field theory limit.

For $N<\infty$ fixed, let $V=\mathbb R^{2N}$ and let $g$ denote the standard Euclidean metric on $V$. Fix an orthonormal basis $\lbrace e_1,e_2,\ldots,e_{2N} \rbrace$ for $V$ and consider the orthogonal complex structure $J$ defined on basis vectors by $J e_k:= e_{N+k}$, $J e_{N+k}:=-e_k$ $(k=1,\ldots, N)$. The 1-particle Hilbert space  corresponding to the Fock representation (\ref{eq:aj})  of the CAR algebra
$\pazocal A_{\mbox{{\tiny CAR}}}(V_J,\langle\cdot,\cdot\rangle_J)$, which is just $(V_J,\langle\cdot,\cdot\rangle_J)$, will be denoted here with
$(\pazocal{H},\braket{\cdot,\cdot})$. Using the convention
\begin{equation}
\label{eq:a_ks}
  a_k \equiv a_J(e_k),\;\; a_k^{\dagger}\equiv a_J^{\dagger}(e_k), \quad k=1,\ldots,N,
\end{equation}
for creation (annihilation) operators, we may recover the basis vectors $e_k$ from the vacuum $|0_J\rangle$ through
\begin{equation}
\label{eq:correspondence}
e_k= a_k^\dagger |0_J\rangle, \quad k=1,\ldots,N.
\end{equation}

We will regard the Hamiltonian (\ref{eq:quadratic-H}) as being written in terms of the fermionic operators (\ref{eq:a_ks}). The idea behind the construction of the self-dual CAR algebra is that, since the ground state of $H$ will in general be different from $|0_J\rangle$, it is more convenient to treat both creation and annihilation operators in a symmetric way. In view of the correspondence (\ref{eq:correspondence}), one possibility is to
consider the ``bras" $\tilde e_k = \langle 0_J|a_k\in \pazocal H^*$, where $\lbrace \tilde e_k\rbrace_k$ is the basis dual to $\lbrace e_k\rbrace_k$. It is then natural to regard (\ref{eq:q-form}) as a quadratic form on $\pazocal H\oplus \pazocal H^*$. But in order not to lose track of the algebra of observables, the self-dual formalism proposes to start anew  by considering a complex Hilbert space $(\pazocal K,\langle\cdot ,\cdot\rangle_{\pazocal K})$
together with a conjugation $\Gamma$, that is, an anti-unitary operator in $\pazocal K$ such that $\Gamma^2=1$. The \textit{self-dual} CAR algebra
$\pazocal A^{\mathrm{ sd}}_{\mbox{\tiny CAR}}(\pazocal K,\Gamma)$ is then defined as the  $*$-algebra  generated by the identity and  operators $B^{\dagger}(u)$, $B(v)$, subject to the following relations ($u, v\in\pazocal K;\; \lambda,\mu \in \mathbb C$):
\begin{align}
&\{B(u),B^\dagger(v)\}=\braket{u,v}_{\pazocal K},\label{eq:sd1}\\
&B^{\dagger}(\lambda u + \mu v)= \lambda B^{\dagger}(u)+ \mu B^{\dagger}(v),\label{eq:sd2}\\
&B^{\dagger}(u)=B(\Gamma u).\label{eq:sd3}
\end{align}
Although these relations resemble the usual fermionic anti-commutation relations, they
are different. In fact, $\Gamma^2=1$, together with (\ref{eq:sd1}) and (\ref{eq:sd3}), implies
$\lbrace B(u), B(v) \rbrace = \langle u,\Gamma v \rangle_{\pazocal K}$.

A first important remark is that it is possible to construct an isomorphism between the self-dual CAR algebra $\pazocal A^{\mathrm{ sd}}_{\mbox{\tiny CAR}}(\pazocal K,\Gamma)$, and the CAR algebra associated to a certain subspace of $\pazocal K$.
In fact,  if $\pazocal K$ is either even or infinite dimensional,  it is possible~\cite{Araki1968} to construct a projection operator  $E$ with the following property:
\begin{equation}
\label{eq:Gamma-E}
\Gamma\, E\, \Gamma=1-E.
\end{equation}
A projection operator satisfying (\ref{eq:Gamma-E}) is called a \emph{basis projection}~\cite{Araki1968}.
It follows that there is an algebra isomorphism $\pazocal A^{\mathrm{ sd}}_{\mbox{\tiny CAR}}(\pazocal K,\Gamma)\cong \pazocal A_{\mbox{{\tiny CAR}}} (E\pazocal K)$, defined on generators by
\begin{eqnarray}
\label{eq:psi}
\psi:\pazocal A^{\mathrm{ sd}}_{\mbox{\tiny CAR}}(\pazocal K,\Gamma) & \longrightarrow & \pazocal A_{\mbox{{\tiny CAR}}} (E\pazocal K)\nonumber \\
B(u)\;\;\;\; & \longmapsto & a(Eu)+a^{\dagger}(\Gamma(1-E)u).\qquad\label{eq:Isom-Psi}
\end{eqnarray}
Using the fact that $E$ is an orthogonal projection and that $\Gamma$ is anti-unitary, one checks that $\psi$ preserves the algebraic structure:
\begin{eqnarray}
  \lbrace \psi(B(u)), \psi (B(v)) \rbrace &=& \langle u,v \rangle_{\pazocal K}\nonumber\\
   &=& \psi (\lbrace B(u), B(v) \rbrace).\quad
\end{eqnarray}
In spite of the fact that $E\pazocal K$ is a (proper) subspace of $\pazocal K$, the algebraic structures they give rise to, are completely equivalent. Notice also how the self-dual algebra allows us to codify both creation and annihilation operators using a single type of operator. According to the isomorphism  (\ref{eq:Isom-Psi}), any operator of the form $B(Eu)$ can be regarded as a fermionic \emph{annihilation} operator, whereas (because of (\ref{eq:Gamma-E})) $B(\Gamma E u)$ can be regarded as a fermionic \emph{creation} operator.

\subsection{Lifting of the Hilbert space redundance and $\mathbb Z_2$-index} Of greater relevance for our purposes is the inverse map to (\ref{eq:psi} (which, to a given CAR algebra $\pazocal A_{\mbox{{\tiny CAR}}} (\pazocal H)$ associates a self-dual CAR algebra
$\pazocal A^{\mathrm{ sd}}_{\mbox{\tiny CAR}}(\pazocal K,\Gamma)$. In order for these two algebras to be related, the Hilbert space $\pazocal H$ should, somehow, be the image of the projection on a bigger Hilbert space $\pazocal K$. But since in this case our initial data is given only by $\pazocal H$, the construction of $\pazocal K$ must  involve some enlarging of $\pazocal H$. As we now show, this corresponds to the usual ``doubling'' of the Hilbert space.   Consider, then, a Hilbert space $(\pazocal H,\langle \cdot,\cdot\rangle)$.
%
%
Define now  $\pazocal K:=\pazocal{H}\oplus\pazocal{H}$ and introduce the projection operator $E$ on $\pazocal K$  given by $E(x,y)=(x,0)$, so that
$\pazocal H= E \pazocal K$. Choose a complex conjugation $T$ on $\pazocal{H}$ and define for $(u,v)\in \pazocal K$ a conjugation $\Gamma(u,v)\coloneqq(Tv,Tu)$. Then we obtain an isomorphism $\pazocal A_{\mbox{{\tiny CAR}}} (\pazocal H)\cong \pazocal A^{\mathrm{ sd}}_{\mbox{\tiny CAR}}(\pazocal H \oplus\pazocal H,\Gamma)$, defined on generators by
\begin{eqnarray}
\label{eq:phi}
\phi: \pazocal A_{\mbox{{\tiny CAR}}} (\pazocal H)  & \longrightarrow & \pazocal A^{\mathrm{ sd}}_{\mbox{\tiny CAR}}(\pazocal H \oplus\pazocal H,\Gamma) \nonumber\\
a(u)& \longmapsto & B^{\dagger}((0,Tu))\equiv B((u,0))\quad
\end{eqnarray}
that defines an algebra isomorphism which, in the case $\pazocal K=\pazocal H\oplus \pazocal H$, is the inverse of the map $\psi$ defined in (\ref{eq:psi}).

The isomorphism (\ref{eq:psi}) (along with its  inverse (\ref{eq:phi})) shows in an explicit way why the doubling of the Hilbert space appearing in the diagonalization of quadratic Hamiltonians like (\ref{eq:q-form}) does not introduce any kind of redundance. In fact, whereas it is natural to consider the space $\pazocal H \oplus\pazocal H$ as a convenient mathematical step in bringing the Hamiltonian to diagonal form, the corresponding algebra of observables is the self-dual algebra $\pazocal A^{\mathrm{ sd}}_{\mbox{\tiny CAR}}(\pazocal H \oplus\pazocal H,\Gamma)$ which is completely equivalent to the original fermionic algebra $\pazocal A_{\mbox{{\tiny CAR}}} (\pazocal H)$. As the latter only contains information about $\pazocal H$, this proves our statement.

We now establish a connection between the self-dual formalism and the one presented in section \ref{S:2}. For this purpose consider the \textit{conjugate Hilbert space} $\overline{\pazocal{H}}$. It has the same underlying set as $\pazocal{H}$ but with  scalar multiplication given by $\lambda$$\cdot$$x\coloneqq \overline{\lambda}x$ for $x\in\overline{\pazocal{H}}$ and $\lambda\in\mathbb{C}$. We also have $\braket{x,y}_{\overline{\pazocal{H}}}\coloneqq{\overline{\braket{x,y}}}=\braket{y,x}$. Set now $\pazocal K\coloneqq\pazocal{H}\oplus\overline{\pazocal{H}}$ with $x\in \pazocal K$  written as $x=(x_{1},x_{2})$. The inner product is defined for $x,y\in \pazocal K$ by $\braket{x,y}_{\pazocal K}\coloneqq\braket{x_{1},y_{1}}+\braket{x_{2},y_{2}}_{\overline{\pazocal{H}}}$. Consider also the projection  $Px\coloneqq (x_{1},0)$ and the complex conjugation
\begin{equation}
\label{eq:complex-conjugation}
\Gamma(x)\coloneqq(x_{2},x_{1}).
\end{equation}
This is clearly a conjugation because $\Gamma(\lambda x)=\overline{\lambda}\,\Gamma (x)$. The operator $P$ is a basis projection on $\pazocal K$ with respect to this conjugation.
Consider now the \textit{real subspace} $\mathrm{Re}_{\Gamma}\pazocal K\coloneqq\{x\in \pazocal K:\Gamma (x)=x\}$. An element  $x\in\mathrm{Re}_{\Gamma}\pazocal K$ must be of the form $x=(x_{1},x_{1})$. It follows that  for $x,y\in\mathrm{Re}_{\Gamma}\pazocal K$ we have $\braket{x,y}_{\pazocal K}=2\mathrm {Re}\braket{x_{1},y_{1}}$. Therefore, the generators  $\pi_{P}(x)\coloneqq a^{\dagger}(Px)+a( P\Gamma x)$ ($x\in \mathrm{Re}_{\Gamma}\pazocal K $) match exactly
 the Clifford generators (\ref{eq:Clifford-generators}). Furthermore, for arbitrary $x\in \pazocal K$, we readily check  that $\pi_{P}(x)=\psi(B^{\dagger}(x))$, with $\psi$ defined as in (\ref{eq:psi}), with $E=P$.

 If we now put $V=\mathrm{Re}_{\Gamma}\pazocal K$ and  $J=i(2P-1)$, $g(u,v)= \mathrm {Re}\braket{u,v}$, we obtain an isomorphism $(V_{J},\braket{\cdot,\cdot}_{J})\cong (\pazocal H, \braket{\cdot,\cdot})$. On the other hand, starting from a triple $(V,g,J)$ as in section \ref{S:2}, and making use of the fact that $W_{-J}\cong \overline{W}_J$ we obtain, for $\pazocal K = W_J\oplus \overline{W}_J$ and $\Gamma$ as above, an isomorphism $\mathrm{Re}_\Gamma \pazocal K \cong V$.
We can summarize our discussion as follows: Let $(V,g,J)$ as in section \ref{S:2}. Let $\pazocal H$ be the complex Hilbert space $V_J$, with scalar product given by (\ref{eq:<,>_J}). Endow $\pazocal H\oplus \overline{\pazocal H}$ with the complex conjugation (\ref{eq:complex-conjugation}). Then  we have the following equivalences:
 \begin{equation}
 \label{eq:equiv}
 \mathbb{C}\ell(V) \cong
 \pazocal A_{\mbox{{\tiny CAR}}} (\pazocal H)  \cong  \pazocal A^{\mathrm{ sd}}_{\mbox{\tiny CAR}}(\pazocal H \oplus\overline{\pazocal H},\Gamma).
 \end{equation}
The last equality in (\ref{eq:equiv}) not only confirms that there is no redundancy in the description of the system, but also provides a direct link between (i) the Bogoliubov transformation that diagonalizes $H$ (ii) the corresponding orthogonal complex structure  and (iii) the topological $\mathbb Z_2$-index. In fact,
 let now $h\in O(V,g)$ be such that the Hamiltonian
(\ref{eq:q-form}) becomes diagonal in the fermionic operators
\begin{equation}\label{eq:Bogoliubov}
  c (v) = a(p_h) +a^\dagger (q_hv),
\end{equation}
where 
\begin{eqnarray}
p_h&=&\frac{1}{2}(h-JhJ),\label{eq:p_h}\\
q_h&=&\frac{1}{2}(h+JhJ), \label{eq:q_h}
\end{eqnarray} 
are the linear/antilinear transformations giving rise to the corresponding Bogoliubov transformation. Then, the ground state of $H$ is characterized by the vacuum condition (\ref{eq:vacuum-condition}) corresponding to the complex structure $J_h\coloneqq hJh^{-1}$. Furthermore, the map~\cite{Gracia-Bond'ia2001}
\begin{eqnarray}\label{eq:index}
  \mathrm{index}: \pazocal J &\longrightarrow & \mathbb Z_2\nonumber\\
  J_h & \longmapsto &  (-1)^{\frac{1}{2}\mathrm{dim}\ker (J +J_h)}
\end{eqnarray}
gives exactly the topological $\mathbb Z_2$-index (Pfaffian invariant). In the next section we illustrate this assertion  with explicit examples.

\section{Examples}
\label{sec:examples}
\subsection{Two-site chain}\label{example1} Let  $V=\mathbb R^4$ with   $g_{\mbox{\tiny E}}(\cdot,\cdot)$ the standard Euclidean metric. For
$e_1,\ldots,e_4$ the standard basis vectors, introduce the  following  complex structure:
\begin{equation}
J=\left(
    \begin{array}{cccc}
      0 & 0 & -1 & 0 \\
      0 & 0 &  0 & -1 \\
      1 & 0 &  0 & 0 \\
      0 & 1 &  0 & 0 \\
    \end{array}
  \right)
\end{equation}
Notice that we have $e_3 = Je_1$ and $e_4=Je_2$. Consider now the following
 Hamiltonian (two-site Kitaev chain):
\begin{equation}
H=t (a_1^\dagger a_2 +a_2^\dagger a_1) +\Delta (a_1^\dagger a_2^\dagger-a_1a_2)-2\mu (a_1^\dagger a_1 + a_2^\dagger a_2).
\end{equation}
Introducing parameters
$\alpha=\sqrt{\Delta^2 +4\mu^2}$,  $\beta_{\pm}=\sqrt{( \alpha\pm \Delta)/(2\alpha)}$
and  $\sigma = \mathrm{sgn}(\alpha-t)$, one readily checks that the Bogoliubov transformation that diagonalizes $H$ is induced by the orthogonal transformation
\begin{equation}
h= \left(
     \begin{array}{cc}
       \Phi & 0 \\
       0 & \Psi \\
     \end{array}
   \right),
\end{equation}
where
\begin{equation}
\Phi=\left(
       \begin{array}{cc}
         \beta_+ & \beta_- \\
         -\beta_- & \beta_+ \\
       \end{array}
     \right),\qquad
\Psi=\left(
       \begin{array}{cc}
         \sigma \beta_- & \sigma \beta_+ \\
         -\beta_+ & \beta_- \\
       \end{array}
     \right).
\end{equation}
For the real maps $p_h,q_h:V\rightarrow V$ (cf. (\ref{eq:p_h}) and (\ref{eq:q_h})), expressed in block form, we find:
\begin{equation}
  p_h =\left(
         \begin{array}{cc}
           g & 0 \\
           0 & g \\
         \end{array}
       \right), \quad
q_h =\left(
         \begin{array}{cc}
           f & 0 \\
           0 & -f \\
         \end{array}
       \right), \quad
\end{equation}
where $g=(1/2)(\Phi+\Psi)$ and  $f=(1/2)(\Phi-\Psi)$ (cf. \cite{Reyes-Lega2016}).
For the  orthogonal complex structure we obtain
  \begin{equation}
    J_h= hJh^\intercal=\frac{1}{\sqrt{\Delta^2 +4\mu^2}}\left(
                                  \begin{array}{cccc}
                                    0 & 0 & -2\sigma\mu & \Delta \\
                                    0 & 0 & -\sigma \Delta & -2\mu \\
                                   2\sigma\mu & \sigma \Delta & 0 & 0 \\
                                    -\Delta & 2\mu & 0 & 0 \\
                                  \end{array}
                                \right).
  \end{equation}
Finally, the $\mathbb Z_2$-index is given by
  \begin{equation}\label{eq:theindex}
  \mathrm{index}(h) := (-1)^{\frac{1}{2}\mathrm{dim}\ker (J +J_h)} = \det h=\sigma.
  \end{equation}
The index defines two regions on the $t$-$\mu$ plane according to the value of $\sigma$. The boundary between these two regions is determined by the condition $\alpha=t$, which can also be written as 
\begin{equation}
\Delta^2 = t^2 -4\mu^2.
\end{equation}
The result is displayed in figure \ref{fig:subfig4}, where we have taken $\Delta = 1$. The shaded region corresponds to $\sigma=-1$ and the unshaded to $\sigma=1$. 
\subsection{$N$-site Kitaev chain }
Finally we  turn our attention to Majorana fermions. A fermionic operator $f$ can be regarded as a superposition of two Majorana fermions, obtained from a splitting into real and imaginary parts:
$f=\gamma^{(A)}+i\gamma^{(B)}$,
with $\gamma^{(A)}$ and $\gamma^{(B)}$ self-adjoint. We can describe Majorana fermions in terms of orthogonal complex structures as follows. Starting with a triple $(V,g,J)$
as above, we describe fermions as elements of  $\pazocal A_{\mbox{{\tiny CAR}}}(V_J,\langle\cdot,\cdot\rangle_J)$. We want to split the generators as
$a_J(v) = \gamma^{(A)}(v) +i\gamma^{(B)}(v)$. Taking into account that $a_J(v)$ depends anti-linearly on $v$ (see equation (\ref{eq:identity2})), we obtain the following characterization of the Majorana operators, in terms of the Clifford algebra generators
\begin{equation}
  \gamma^{(A)}(v) = \frac{1}{2}\pi_J(v), \quad
   \gamma^{(B)}(v) = \frac{1}{2}\pi_J(Jv).
\end{equation}
We see that the orthogonal complex structure allows us to identify the Majorana modes.
Consider the generalization of example \ref{example1} to the case of a Kitaev chain with $N$ sites:
\begin{equation}\label{eq:Kitaevchain}
H=\sum_{i=l}^N t (a_i^\dagger a_{i+1} +a_{i+1}^\dagger a_i) +\Delta (a_i^\dagger a_{i+1}^\dagger-a_ia_{i+1})-2\mu a_i^\dagger a_i.
\end{equation}

The matrices $\Phi$ and $\Psi$ will now be $N\times N$ matrices. For open boundary conditions, the exact solution involves the solution of a trascendental equation~\cite{Reyes-Lega2016}. The Majorana edge modes can nevertheless still be obtained numerically, and are given by
\begin{equation}
\gamma_{\mbox{\tiny edge}}^{(A)} = \sum_{i=1}^N\Phi_{1i}\gamma_i^{(A)},\quad
\gamma_{\mbox{\tiny edge}}^{(B)} = \sum_{i=1}^N\Psi_{1i}\gamma_i^{(B)}.
\end{equation}
\begin{figure}
\centering
\subfloat[$N=4,r=0$]{
\includegraphics[width=0.33\textwidth]{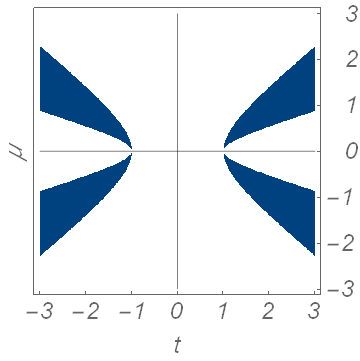}
\label{fig:subfig1}}
\subfloat[$N=4,r=0.1$]{
\includegraphics[width=0.33\textwidth]{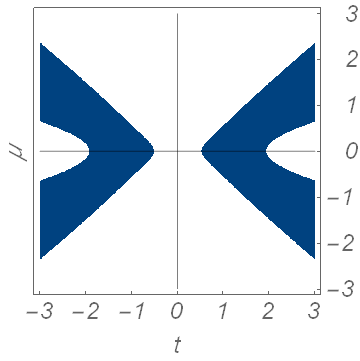}
\label{fig:subfig2}}
\subfloat[$N=4,r=0.2$]{
\includegraphics[width=0.33\textwidth]{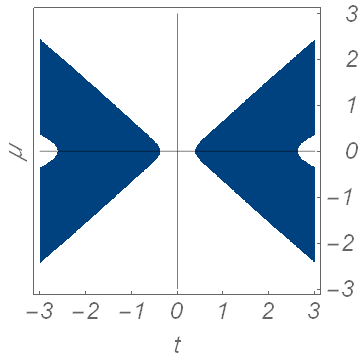}
\label{fig:subfig3}}

\subfloat[$N=2,r=0$ ]{
\includegraphics[width=0.33\textwidth]{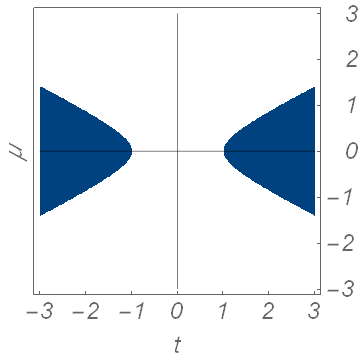}
\label{fig:subfig4}}
\subfloat[$N=6,r=0$ ]{
\includegraphics[width=0.33\textwidth]{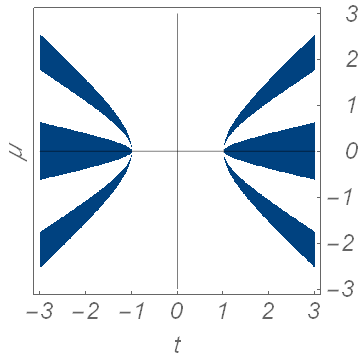}
\label{fig:subfig5}}
\subfloat[$N=8,r=0$ ]{
\includegraphics[width=0.33\textwidth]{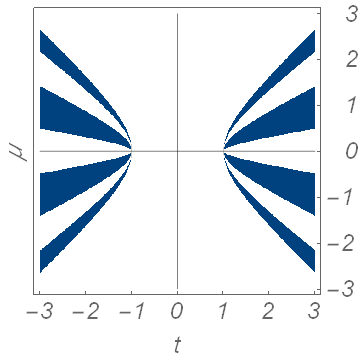}
\label{fig:subfig6}}

\caption{The shaded regions correspond to nontrivial values of the $\mathbb Z_2$-index associated to the Kitaev chain Hamiltonian eq.(\ref{eq:Kitaevchain}), with $\Delta=1$. $N$ denotes the number of sites in the chain and $r$ is a parameter that allows us to interpolate between open boundary conditions ($r=0$) and periodic ones ($r=1$).}
\label{fig}
\end{figure}

The $\mathbb Z_2$-invariant can still be computed using formula (\ref{eq:index}). This holds even in the limit $N\rightarrow \infty$, because $J_h-J$ is a Hilbert-Schmidt operator. Notice that in this approach the $\mathbb Z_2$-invariant is computed in the same way (as the index of a Fredholm operator) irrespective of whether we are using periodic or open boundary conditions. However, it is important to remark that the correspondence between the non-trivial value of the invariant and the actual appearance of edge modes only occurs if we use periodic boundary conditions. This is illustrated in figure \ref{fig}, where the shaded regions in the $\mu$-$t$ plane correspond to the value $-1$ for the  $\mathbb Z_2$-index. Introducing a real parameter $r\in [0,1]$ such that $r=0$ corresponds to open boundary conditions and $r=1$ to periodic boundary conditions, we see from figures \ref{fig:subfig1}, \ref{fig:subfig2} and \ref{fig:subfig3} that the region where the  $\mathbb Z_2$-index takes the value $-1$ will only coincide with the region in parameter space supporting edge states, when periodic boundary conditions ($r=1$) are being used. Otherwise, the index provides information on the parity of the ground state. It is interesting, nevertheless, that for open boundary conditions the index formula  (\ref{eq:index}) can still be applied, as depicted in figures \ref{fig:subfig4}, \ref{fig:subfig5} and \ref{fig:subfig6}.
 %
\section{Discussion}\label{S:5}
The approach presented here, which is based on the use of orthogonal complex structures for the description of fermionic systems, has allowed us to give a unified account of several aspects of relevance in the context of topological phases of matter. We have shown that, once correctly incorporated,  an orthogonal complex structure produces a reduction in the dimension of the Hilbert space, accounting for the ``redundancy'' usually discussed in the context of Hamiltonians of the BdG form. This has also been highlighted by constructing an explicit isomorphism between the fermionic CAR algebra and a (variation) of Araki's self-dual CAR algebra. Furthermore, we have shown in an explicit way how the $\mathbb Z_2$-invariant is related to the complex structure.    In particular, our formalism allows for a direct computation of the $\mathbb Z_2$-invariant independently of the boundary conditions chosen. This might be of help in the search for a rigorous proof of the bulk-boundary correspondence for cases where it still remains at the level of a conjecture~\cite{Bourne2016}. An additional advantage of our approach is that it allows for the inclusion of disorder and thus the topological invariance can be explicitly tested under more realistic conditions. We hope to report on these issues in future work.

\section*{Acknowledgments}

The authors would like to thank A.P. Balachandran for fruitful discussions during different stages of this work.
Financial support from the Faculty of Science and the Vice
Rectorate for Research of Universidad de los Andes, through project No.
P13.700022.005, is gratefully acknowledged.


%

\end{document}